# Neutron calibrations in dark matter searches: the ANAIS-112 case

**Tamara Pardo,**[a,b,*] **Julio Amaré,**[a,b] **Jaime Apilluelo,**[a,b] **Susana Cebrián,**[a,b] **David Cintas,**[a,b] **Iván Coarasa,**[a,b] **Eduardo García,**[a,b] **María Martínez,**[a,b] **Miguel Ángel Oliván,**[a,b,c] **Ysrael Ortigoza,**[a,b,d] **Alfonso Ortiz de Solórzano,**[a,b] **Marta Pellicer,**[a] **Jorge Puimedón,**[a,b] **Ana Salinas,**[a,b] **María Luisa Sarsa**[a,b] **and Patricia Villar**[a]

[a]*Centro de Astropartículas y Física de Altas Energías (CAPA), Universidad de Zaragoza*
*Pedro Cerbuna 12, 50009 Zaragoza, Spain*

[b]*Laboratorio Subterráneo de Canfranc*
*Paseo de los Ayerbe, 22880 Canfranc Estación, Huesca, Spain*

[c]*Fundación CIRCE*
*Avenida de Ranillas 3D, 50018 Zaragoza, Spain*

[d]*Escuela Universitaria Politécnica de La Almunia de Doña Godina (EUPLA), Universidad de Zaragoza*
*Calle Mayor 5, La Almunia de Doña Godina, 50100 Zaragoza, Spain*

*E-mail:* tpardo@unizar.es

ANAIS is a direct dark matter detection experiment whose goal is to confirm or refute in a model independent way the positive annual modulation signal claimed by DAMA/LIBRA. Consisting of 112.5 kg of NaI(Tl) scintillators, ANAIS-112 is taking data at the Canfranc Underground Laboratory in Spain since August, 2017. Results corresponding to the analysis of three years of data are compatible with the absence of modulation and incompatible with DAMA/LIBRA. However, testing this signal relies on the knowledge of the scintillation quenching factors (QF), which measure the relative efficiency for the conversion into light of the nuclear recoil energy with respect to the same energy deposited by electrons. Previous measurements of the QF in NaI(Tl) show a large dispersion. Consequently, in order to better understand the response of the ANAIS-112 detectors to nuclear recoils, a specific neutron calibration program has been developed. This program combines two different approaches: on the one hand, QF measurements were carried out in a monoenergetic neutron beam; on the other hand, the study presented here aims at the evaluation of the QF by exposing directly the ANAIS-112 crystals to neutrons from low activity $^{252}$Cf sources, placed outside the lead shielding. Comparison between these onsite neutron measurements and detailed GEANT4 simulations will be presented, confirming that this approach allows testing different QF models.

[*]Speaker

## 1. Introduction

The ANAIS-112 experiment [1, 2] aims to provide a model independent confirmation or refutation of the DAMA/ LIBRA positive annual modulation signal [3] using the same target, 112.5 kg of NaI(Tl), and technique. It is taking data at the Canfranc Underground Laboratory (LSC) in Spain since August 3, 2017. Annual modulation results with three-year exposure are compatible with the absence of modulation and incompatible with DAMA/LIBRA result at $3.3\sigma$ ($2.6\sigma$) in [1-6] keV ([2-6] keV), for a sensitivity of $2.5\sigma$ ($2.7\sigma$) [2], which increases up to $2.9\sigma$ in both energy regions [4] after reanalysis applying machine learning techniques for data filtering [5].

ANAIS and DAMA/LIBRA use the same target material, NaI(Tl), and both experiments are calibrated using X-rays/gamma lines, which release the energy via electron recoils (ER). Consequently, their measurements are presented in terms of electron-equivalent energy and they can be directly compared only if energy is deposited by this channel. This caveat is particularly relevant in the testing of the DAMA/LIBRA result, because scintillation from nuclear recoils (NR) is strongly quenched with respect to the same energy deposited by ER. The quenching factor (QF) is defined as the ratio of the light yield from NR to that from ER depositing the same energy. For many dark matter (DM) candidates, in particular Weakly Interacting Massive Particles, the dominant interaction channel is the elastic scattering off the target nuclei, i.e. energy is released through NR. This leads to the requirement of having a good knowledge of QFs in order to recalibrate in terms of NR energies before any interpretation of the measurements regarding DM particles. However, measurements of the QF of sodium and iodine nuclei in NaI(Tl) do not agree and are affected by a large dispersion (see [6] and the embedded references): constant values have been usually considered, for instance by DAMA/LIBRA (30% and 9% for Na and I recoils, respectively), but dependence with energy is observed in many of the most recent measurements. Two different scenarios could be considered: on the one hand, the QF may indeed be an inherent property of NaI(Tl) and the differences among measurements might be produced by unaccounted systematics in the experimental and analysis procedures; on the other hand, the QF could depend on specific crystal properties, as for example impurities or doping concentration. The former situation implies that comparing experiments using NaI(Tl) is direct, while the latter, that specific calibrations for NR for each crystal batch are required. To address these discrepancies and better understand the response of the ANAIS-112 detectors to different particle energy depositions, a specific neutron calibration program is currently under development.

## 2. The ANAIS-112 neutron calibration program

The ANAIS-112 neutron calibration program combines two different approaches. QF measurements were initially conducted at TUNL (Duke University, North Carolina, US) using a monochromatic neutron beam. Five small crystals, including one from the very same ingot as several ANAIS-112 modules, were measured using the same setup and analysis. Despite differences in their starting powder quality, they all shared a similar growth procedure and Tl content. Subsequently, neutron calibrations of the ANAIS-112 modules onsite have been performed, starting in 2021, using $^{252}$Cf sources of low activity at LSC. Each approach comes with its own drawback, but they are complementary and should be consistent. While measurements at TUNL use different crys-tals, PMTs and DAQ system than ANAIS-112, those onsite are dominated by multiple scattering, relying then strongly its interpretation on the Monte Carlo (MC) modelling.

The measurement procedure at TUNL and subsequent data analysis is detailed in [6]. A very relevant output is that the procedure for the energy calibration of the NaI(Tl) signal is highlighted as the main systematic affecting QF estimation. Different calibration strategies were followed, both proportional and non-proportional, using $^{133}$Ba calibration data and the 57.6 keV inelastic peak from $^{127}$I. Measurements covered a range from 10 to 80 keV$_{nr}$ sodium NR energies, implying a ROI in electron-equivalent energy below 30 keV. Extrapolating down the energy calibration is mandatory, and then, these measurements were not conclusive. Two different Na-QF modelling are derived from the data depending on the energy calibration procedure chosen: (21.2±0.8)% energy-independent (for non-proportional energy calibration using 6.6, 30.9 and 35.1 keV lines from $^{133}$Ba), and an energy-dependent, from 13% to 25% in the range from 0 to 100 keV, and constant at 25% above 100 keV (for the proportional calibration with 57.6 keV as reference). Regarding the I-QF, it could be estimated as (6.0±2.2)% at 14 keV.

Neutron calibrations onsite provide a clean population of bulk scintillation events, and are crucial for cross-checking QF measurements at TUNL. Bulk scintillation events in the NaI crystals at low energy (<10 keV) are selected by pulse shape analysis, as detailed in [1]. In this study, a non-proportional electron-equivalent energy calibration is used. It considers only bulk scintillation events from internal background lines ($^{22}$Na and $^{40}$K) well identified by coincidence with a high energy gamma in a second module, and both, electron and gamma contributions following neutron interactions (31 keV and 57 keV from $^{128}$I decay and inelastic scattering in $^{127}$I, respectively).

The interpretation of these measurements relies strongly on the availability of a robust MC modelling of the full ANAIS-112 setup, performed using GEANT4, because the measured rates are dominated by multiple elastic scattering on Na and I nuclei. Different modelling for the QF can be introduced in this simulation to convert the energy deposited into electron equivalent energy before comparing with measurements. Results are shown in Figure 1. Regarding the QFs of DAMA/LIBRA (green), a clear discrepancy between simulation and measurement (black) is observed. Nevertheless, when considering the QF resulting from the measurements at TUNL, a better agreement is found with a constant Na-QF (blue) and a much better one with an energy-dependent Na-QF (red). A constant value of 6% is applied for I-QF, and 1$\sigma$ uncertainty bands are drawn. Indeed, the energy-dependent Na-QF shows a good qualitative agreement, being favored when compared to constant Na-QF, although some small mismatching is still under study.

## 3. Conclusions

The relative efficiency for the conversion of the NR energy into light in NaI(Tl) is a very important systematic in the testing of DAMA/LIBRA result. With the goal of determining the QF of the ANAIS crystals, dedicated measurements onsite with $^{252}$Cf sources have been performed. A GEANT4 simulation of the whole ANAIS-112 experimental set-up allows to reproduce the experimental measurements for different QF modelling. Comparison between data and simulation shows that energy-dependent Na-QF reproduce satisfactorily the measured energy spectra for single and multiple-hits, and is favoured over constant QF models. This analysis is still ongoing and further studies are required to fully understand the QF of the ANAIS-112 detectors.

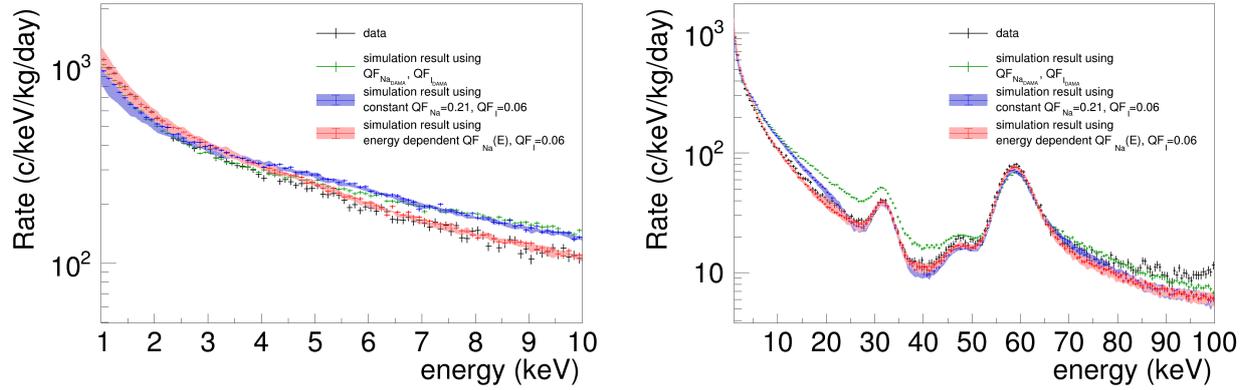

**Figure 1:** Comparison of the total measured energy spectrum for $^{252}$Cf calibrations of the ANAIS-112 experiment (black) with the simulation when different QFs models are considered: DAMA/LIBRA QFs (green), constant Na-QF (blue) or energy dependent Na-QF (red). $1\sigma$ uncertainty bands are shown as blue and red shaded areas, respectively. Horizontal scale is represented in electron-equivalent energies. Left panel: low energy range; right panel: medium energy range.

## Acknowledgments

This work has been financially supported by MCIN/AEI/10.13039/501100011033 under grants PID2022-138357NB-C21 and PID2019-104374GB-I00, the Consolider-Ingenio 2010 Programme under grants MultiDark CSD2009-00064 and CPAN CSD2007-00042, the LSC Consortium, and the Gobierno de Aragón and the European Social Fund (Group in Nuclear and Astroparticle Physics) and funds from European Union NextGenerationEU/PRTR (Planes complementarios, Programa de Astrofísica y Física de Altas Energías). Authors would like to acknowledge the use of Servicio General de Apoyo a la Investigación-SAI, Universidad de Zaragoza and technical support from LSC and GIFNA staff.

## References

[1] J. Amaré et al., *Performance of ANAIS-112 experiment after the first year of data taking*, *Eur. Phys. J. C* **79** (2019) 228 [1812.01472].

[2] J. Amaré et al., *Annual modulation results from three-year exposure of ANAIS-112*, *Phys. Rev. D* **103** (2021) 102005 [2103.01175].

[3] R. Bernabei et al., *The DAMA project: Achievements, implications and perspectives*, *Prog. Part. Nucl. Phys.* **114** (2020) 103810.

[4] I. Coarasa et al., *ANAIS-112: updated results on annual modulation with three-year exposure*, *in these procedings* .

[5] I. Coarasa et al., *Improving ANAIS-112 sensitivity to DAMA/LIBRA signal with machine learning techniques*, *JCAP* **11** (2022) 048 ; erratum *JCAP* **06** (2023) E01 [2209.14113].

[6] D. Cintas, *New strategies to improve the sensitivity of the ANAIS-112 experiment at the Canfranc Underground Laboratory*, PhD thesis, Universidad de Zaragoza (2023) [arXiv:2310.07339].